# Nucleophilic substitution vs elimination reaction of bisulfide ions with substituted methanes: exploration of chiral selectivity by stereodirectional first-principles dynamics and transition state theory

Marcos Vinícius C. S. Rezende[1], Nayara D. Coutinho[2*], Federico Palazzetti[2], Andrea Lombardi[2], and Valter Henrique Carvalho-Silva[1*]

[1]Grupo de Química Teórica e Estrutural de Anápolis, Campus de Ciências Exatas e Tecnológicas. Universidade Estadual de Goiás, Anápolis, Brazil

[2]Dipartimento di Chimica, Biologia e Biotecnologie, Università di Perugia, Perugia, Italy

**\* Correspondence:**

nayaradcoutinho@gmail.com

fatioleg@gmail.com

**Abstract**

Control of molecular orientation is emerging as crucial for the characterization of the stereodynamics of kinetics processes beyond structural stereochemistry. The special role played in chiral discrimination phenomena has been particularly emphasized by Aquilanti and collaborators after their extensive probes of experimental control of molecular alignment and orientation. In this work, the manifestation of the Aquilanti mechanism has been demonstrated for the first time in first-principles molecular dynamics simulations: stationary points characterized on potential energy surfaces have been calculated for the study of chemical reactions occurring between the bisulfide anion HS − and oriented prototypical chiral molecules CHFXY (where X = CH3 or CN and Y = Cl or I). The important reaction channels are those corresponding to bimolecular nucleophilic substitution (SN2) and to bimolecular elimination (E2): their relative role has been assessed and alternative pathways due to the mirror forms of the oriented chiral molecule are revealed by the different reactivity of the two enantiomers of CHFCNI in SN2 reaction.

## Introduction

Chiral selectivity is a topic that involves numerous areas of science, interest for this phenomenon raises from the widespread presence of processes characterized by chiral specificity in nature, and, most of all, from the phenomenon of homochirality of biomolecules, that involves aminoacids and sugars [1][2]. The stereodynamical origin of chiral discrimination opens ample perspectives beyond traditional dichroism studies that make use of circularly polarized radiation [3]. Its relevance spans from the industrial and pharmaceutical area, until astrochemistry [4, 5], culminated with the recent discoveries of hydrogen peroxide [6] and propylene oxide [7] in the molecular clouds of the interstellar medium.

Control of direction, velocity and internal states of molecular beams are basic in chemical reaction kinetics to predict and interpret the mechanism of a collisional process [8]. Evolution of molecular beam techniques permitted to open the way to the study of enantioselective collisional processes, in order to establish the role played by molecular anisotropy [9, 10]. A fundamental step has been taken with the development of molecular alignment and orientation techniques [11]. It is defined alignment a phenomenon of polarization of the rotational degrees of freedom of molecules, for which it is possible to select only determinate directions. Orientation is a stronger condition that permits to select not only the direction, but also the sense. Alignment and orientation techniques can be performed by exploiting the seeding effect in supersonic beams or resorting to the influence of an external field, such as electric fields or laser beams. Here, we report two main techniques developed by our group: the natural alignment and the electrostatic hexapole orientation. The natural alignment technique has been discovered in the middle of the Nineties in Perugia [12] and applied to oxygen [13], nitrogen [14], benzene [15] and ethylene [16] molecules. The principle of this technique is based on the expansion of a supersonic molecular beam seeded in lighter atoms (for example helium), where collisions determine a rotational relaxation of the molecule and, subsequently, an alignment with respect to the main propagation direction of the beam, whose characteristic can varies with the different molecules. A further inspiring addition to the picture of the alignment effect in gaseous flows, is due to Su *et al.,* who have demonstrated that gaseous flows characterized by rotary motions, induced by the action of screw pumps inducing vortices, can generate chiral effects on the chiral molecules seeded in the beam, in the form of a detectable chiral enrichment in a single enantiomer, dependent on the sense of rotation by screw pumps [17–19]. The microscopic nature of such a mechanism must be still interpreted [20]. The electrostatic hexapolar orientation is carried out by an inhomogeneous electrostatic field produced by a hexapole, which induce molecular alignment, followed by a homogenous field, whose effect is to orient the molecules. This technique was for the first time applied to chiral molecules a decade ago on propylene oxide [21, 22] and more recently on 2-butanol [23].

Molecular orientation is fundamental to observe spatial aspects that would be hidden by the averaging effect of the random rotation, on this basis the orientation turns out to be a fundamental requisite for the observation of chiral effects in molecular collisions. It is defined "Aquilanti mechanism" the *mechanism mediated by molecular orientation that permits to discriminate between a pair of enantiomers*. Experiments on oriented molecules have been carried out to investigate photodissociation dynamics of chiral molecules [24–26]. Although the Aquilanti mechanisms has not been yet clarified from an experimental point-of-view, from a theoretical point-of-view it has been demonstrated in elastic collisions simulations [27] of the $H_2O_2$ [28] and $H_2S_2$ [29] chiral molecules with rare gas atoms. Vector correlation permitted also to demonstrate the enantioselective effect in dissociation processes initiated by linearly polarized lasers [30]. In this work, we give for the first time the proof of the role played by molecular orientation in reactive collisional processes. Specifically, we have investigated a bimolecular nucleophilic substitution ($S_N2$) reaction.

Theoretical and experimental studies on $S_N2$ mechanisms have shown the presence an alternative competing channel, the second-order elimination mechanism (E2) [31–36]. These channels are arguably the most fundamental ones in the study of chemical reactions involving organic molecules that permit to establish the stereospecific role for chiral substrates [37]. Theoretical studies in the gas-phase allows the verification of the mechanisms in the absence of solvation effects and, therefore, it becomes possible the intrinsic analysis of the reactivity. Gas-phase studies are important because, in general, the results show that the fundamental characteristics are the same as those identified in the condensed-phase [31]. Experimental characterization of $S_N2$ and E2 pathways is somewhat complicated, since both lead to the same ionic product. Thus the direct probing by mass spectrometry is infeasible [31]. Furthermore, studies for these mechanisms show that in reactive systems, where both bimolecular nucleophilic substitution and bimolecular elimination can occur, new mechanisms beyond the classic collinear back-side attack (Walden inversion configuration) have been revealed and characterized [38, 39].

There are several studies evaluating stationary points on minimum potential energy pathway in classical gas-phase E2 and $S_N2$ channels (see for example [40–42]). The results present a diversity of intrinsic reaction pathways for these processes, such as the double inversion indirect mechanism with proton-transfer for the nucleophilic group. Even with important contribution of the conventional direct mechanism, at low collision energies the indirect mechanism is majority [39, 43–45]. Concerning the competing E2 and SN2 mechanisms, a blend of high-level electronic structure calculations and single-collision experiments on the archetypical $X^-$ + RY reaction, with X and Y denoting F, Cl, Br, $OH^-$ and $CN^-$ groups, confirmed the following points: i) synchronous E2 transition state, instead of nonconcerted mechanism, with preference by anti-E2 configuration; ii) change of the halogen leaving atom (from F to I) causes a decrease in barrier height; iii) $S_N2$ with retention of configuration is a least favourable pathway, Walden inversion configuration pathway being preferred; and iv) abrupt change from backward to dominant forward scattering depending on the alkyl group in the substrate.[36, 46, 47] However, there are scarce studies focused in evaluating the role of the chirality in the E2 and $S_N2$ gas-phase mechanism, highlighting the Rablen study[48] in liquid-phase which elucidates and quantifies the effects on the barrier height of 47 alkyl halide structures, with several chiral substrates. Their results, with rare exceptions, does not reveal significant differences among E2 and $S_N2$ barrier energies.

**Plan of the paper**

Here, we aim at broadening the knowledge of the chiral collisional mechanisms, through extensive theoretical studies of the interaction of nucleophilic species with chiral substrates. As a preparatory work on model systems with interest for astrochemistry and biochemistry, here our calculations evaluate the stationary points responsible for the thermodynamic and kinetic control in $HS^-$ + CHFXY (X = $CH_3$ and CN; Y = Cl and I) reaction. Bisulfide ($HS^-$) was chosen considering its strong nucleophiolic strength; and CHFXY is a prototypical chiral substrate. Preliminary *ab initio* molecular dynamics trajectories are also are presented for $HS^-$ + CHFCNI. The article is organized as follows: section 2 presents a brief summary of the computational details employed in our calculations an of the *deformed* Transition-State Theory formulation (*d*-TST); we present and discuss our results in section 3; we conclude the work with a summary of our achievements in section 4.

# Computational procedure

**Electronic Structure Calculations of stationary points**

The geometry optimizations and energies of reactants, products and transition states have been calculated at a ωB97XD/aug-cc-pVDZ level of theory. Higher level coupled cluster theory with triple excitations treated perturbatively CCSD(T) has been used to further refine the energies. The small-core pseudopotential (LANL2DZ) has been used to describe valence electrons of the iodine atom. The stationary points were characterized by analytic harmonic frequency calculations: the absence or existence of an imaginary frequency indicates that the optimized structures are local minima or transition states, respectively. The zero-point vibrational energy (ZPVE) contributions were considered in the calculation of the energy barriers. Thermodynamic (, and ) and kinetic (, and ) energy parameters have been calculated as the difference between the energy of the products and reactants and the difference between the energy of the transition-state and that of the reactant structures, respectively. All calculations have been performed using the Gaussian09 package[49].

**Reaction Rate Theory**

The reaction rate constant ($k$) has been calculated by *deformed*-Transition state theory($d$-TST)[50, 51] using the Transitivity code (www.vhcsgroup.com/transitivity). For a general bimolecular reaction such as $R_1 + R_2 \rightarrow TS^\dagger \rightarrow$ PRODUCTS, it is necessary to compute the partition functions $Q_1$, $Q_2$ and $Q^\dagger$ of $R_1$, $R_2$ and of the transition state, respectively. At absolute temperature $T$, the rate constant is given by:

$$k(T) = \frac{k_B T}{h} \frac{Q^\dagger}{Q_1 Q_2} \left(1 - d \frac{\varepsilon^\ddagger}{RT}\right)^{1/d}, \qquad d = -\frac{1}{3}\left(\frac{h\nu^\ddagger}{2\varepsilon^\ddagger}\right)^2, \tag{1}$$

where $k_B$ is the Boltzmann's constant, $h$ is the Planck's constant and $\varepsilon^\ddagger$ is the effective height of the energy barrier, given by $\varepsilon^\ddagger = V + \varepsilon_{ZPE}$, where $\varepsilon_{ZPE}$ is the harmonic zero-point energy correction and $V$ is the height of the potential energy barrier. This equation uniformly covers the range from classical to moderate tunneling regimes [52, 53].

**Molecular dynamics**

A series of *ab initio* Born-Oppenheimer molecular dynamics (BOMD) simulations have been carried out using the CPMD 3.17.1 package [54]. The system has been modelled using a periodically repeated cubic cell of side-length 13 Å, containing the two HS⁻ and CHFCNI molecules. Simulations have been performed for the two enantiomeric forms R and S, starting from the same initial configurations. The electronic structure has been treated within the generalized gradient approximation to density functional theory, using the Perdew–Burke–Ernzerhof (PBE) exchange–correlation functional.[55, 56]

Vanderbilt ultrasoft pseudopotentials have been employed to represent core-valence electron interactions [57]. A plane-wave basis set has been used to expand the valence electronic wave functions with an energy cutoff of 25 Ry. The equations of motion have been integrated using a time step of 5 a.u (0.121 fs) for a total time of 1.8 ps. The simulations have been conducted in a NVT ensemble at temperature of 200 K controlled using a Nosé-Hoover thermostat.[58]

# Results and discussion

In Figure 1, we report the transition state molecular structures of the six reactions studied in the R and S configuration at $\omega$B97XD/aug-cc-pVDZ level of calculation. There are not experimental parameters to be compared with the theoretical substrate structure, however the agreement of the geometric parameters obtained for the HS⁻ and H₂S molecules with the corresponding experimental results[59] confirms the quality of the calculation level used to describe the reaction. As reported by Yang *et al.*[46], the anti-E2, [(1) and (2) reactions], and back-side S$_N$2, [(3) to (6) reactions] channels are the most favored ones, thus only these mechanisms have been considered in this work. Anti-E2 and back-side S$_N$2 channels present a synchronous transition state and in each specific channel there are very similar geometric parameters for the transition-state structure, except in the C − Y bond. Noteworthy in the S$_N$2 channel there is the variation of about 10 degrees in the S-$\hat{C}$-Y angle by changing the X group, reflecting a steric effect of the methyl group. Comparing the transition-state geometric parameters obtained by us with similar channels studied in Ref. [46], there are specific differences in breaking and formation of bonds, clearly due to the difference between the ligands, however the molecular configurations are generally very similar in both cases.

In order to rationalize the influence of the different substituents (X=CH₃ and CN) and leaving group (Y=Cl and I) on the relative barriers for S$_N$2 and E2 channels, different X-Y combination have been studied as shown in Figure 1. Using the composite CCSD(T)/aug-cc-pVDZ//$\omega$B97XD/aug-cc-pVDZ level of calculation, the thermodynamic ($\Delta E$, $\Delta H$ and $\Delta G$) and kinetic ($\varepsilon^{\ddagger}$, $\Delta H^{\ddagger}$ and $\Delta G^{\ddagger}$) parameters with zero-point energy were obtained and are listed in Table 1. As expected, no energy differences have been found between the enantiomeric forms, both in S$_N$2 and E2 reaction mechanisms. All S$_N$2 channels are exothermic processes, while the E2 channel (1), with chloride as leaving group, is thermoneutral ($\Delta_r H = 0$ kcal/mol) and the E2 (2), with iodine as leaving group, is endothermic ($\Delta_r H = 13.3$ kcal/mol). We have also calculated the $\Delta_r G$ to get insights into the spontaneity of the reactions. Only reaction (2) can be considered thermodynamically unfavorable ($\Delta_r G > 0$), while the S$_N$2 channel with Cl as leaving group (Reactions (3) and (5)) are the most spontaneous. From the kinetic ~~energetic~~ parameters, it is possible to observe the highest barrier height in the elimination channels, around 10 kcal/mol for chloride as leaving group and 13 kcal/mol for iodine as leaving group. The barrier height decreases significantly for nucleophilic substitutions with the same reactants studied in the elimination, with values around 5 and 6 kcal/mol for chloride and iodine leaving group, respectively. In agreement with literature, back-side $S_N2$ channels with X= CN and Y = Cl and I present negative activation energies of *ca.* -5 kcal/mol, which indicates a favorable pathway prevailing over the other channels. This minimum energy pathway shape is characterized as a chemical process limited by molecular reorientation requirements[53, 60–62], which at low temperatures let incident reactants to reorient and find propitious alignment leading to reactions~~,~~ while, at high temperatures, speed hinders adjustment of directionality and delays the reaction.

The *deformed*-transition state theory (d-TST) has been employed to calculate the kinetic rate constant for $HS^-$ + $CHFCH_3Y$ (Y = Cl and I) reactions in a wide range of temperature (200-4000K) at CCSD(T)/aug-cc-pVDZ//$\omega$B97XD/aug-cc-pVDZ level of calculation (Figure 2). No differences have been found in both enantiomeric forms. The branching ratios between E2 and $S_N2$ channels have been calculated, being defined as the ratios of the rate constant of specific channel to the overall rate constant for each channel, $\frac{k_{channel}}{k_{E2}+k_{S_N2}}$. These branching ratios are used to emphasize the importance of a given channel over the other ones and can indicate the preference for a channel at different temperatures. The $S_N2$ channel clearly is th epreferred one at low temperatures, while, as temperature increases, the contribution of the E2 channel becomes more important until both paths contribute equally, independently of the leaving group.

The stationary electronic structure calculations are unable to differentiate the role of the chiral discrimination in title reactions studied, in agreement with previous studies that suggest the dynamic and orientational effects in chiral selectivity. However, the stationary electronic calculations suggest back-side $S_N2$ channels with X= CN and Y = I as an archetypical chiral channel to be evaluated by *ab initio* molecular dynamics, since the negative activation energy ensure high reactivity at low temperature. Reactive trajectories have been monitored by following the coordinate $s$, defined as the difference between the length of the breaking bond ($r_1 = C - I$) and the length of the forming bond ($r_2 = C - SH$). Negative values of $s$ correspond to the configuration of reagents, while positive values to those of products. As a characteristic signature of completion of the reactive event we can consider the instant at which $s = 0$. Figure 3 reports $s$ as a function of time. Both reactions lead to $S_N2$ products, the R enantiomer decreases its reactivity, since a longer reaction time is required. These preliminary simulations seem to suggest that stereodynamic effect play a role in kinetic of these forms, however statistical trajectories are being performed to support the orientational alignment and will be presented in a future work.

## Conclusion

The reaction between CHFXY (X = CH3 and CN; Y = Cl and I) and HS has been investigated by Born-Oppenheimer Molecular Dynamics trajectory simulations occurring toward $S_N2$ and E2 mechanisms. Reaction mechanisms have been characterized both a termodynamic and kinetic point-of-view. Stationary electronic structure methods confirmed to be unsuitable to discriminate enantiomers, althoguh they permitted to select a reaction path, the back-side $S_N2$ with X= CN and Y = I, as an ideal candidate to verify the dynamics and orientational effect for the chiral selection, being highly reactive at low temperature.

The role played by molecular orientation in chiral discrimination mechanisms, the Aquilanti mechanism, has been demosntrated for the reaction CHFXY (X = CH3 and CN; Y = Cl and I) and HS. The alternative pathways corresponding to R and S mirror forms are observed by monitoring the reaction coordinate *s* of $S_N2$, corresponding to the difference between the length of the forming and the breaking bonds, resulting in a lower reactivity for the R enantiomer. No quantitative assessment has been obtained but to establish the experimental observability, future work is demanded for a statistically valid sampling of reactive trajectories.

# Acknowledgments

The authors are grateful for the support given by CAPES and CNPq. This research is also supported by the High Performance Computing Center at the Universidade Estadual de Goiás (UEG). Valter H. Carvalho-Silva thanks CNPq for the research funding programs [Universal 01/2016 - Faixa A - 406063/2016-8]. Federico Palazzetti, Nayara D. Coutinho and Andrea Lombardi acknowledge the Italian Ministry for Education, University and Research, MIUR, for financial support: SIR 2014 ''Scientific Independence for young Researchers'' (RBSI14U3VF).

**CAPTIONS**

**MANCANO LE CAPTIONS**

Table 1.

**Figure 1.** Geometries and structural parameters optimized for the anti-E2 and side-back $S_N2$ transition-states of the HS$^-$ + CHFXY (X = CH$_3$ and CN; Y = Cl and I) reaction in R and S configurations at ωB97XD/aug-cc- pVDZ level of calculation. Bond lengths are in angstroms, and bond angles are in degrees.

Figure 2.

Figure 3.

**TABLES AND FIGURES**

## Table 1

| Index | Reaction | | | | | |
|---|---|---|---|---|---|---|
| (1) | | -1.0(-1.0) | 0.0(0.0) | -9.9(-9.9) | 10.1(10.1) | 10.2(10.2) |
| (2) | | 12.5(12.5) | 13.3(13.3) | 3.8(3.8) | 13.1(13.1) | 13.0(13.0) |
| (3) | | -20.5(-20.6) | -20.7(-20.8) | -19.0(-19.1) | 5.0(4.9) | 4.9(4.8) |
| (4) | | -4.8(-4.8) | -5.1(-5.1) | -3.4(-3.4) | 6.4(6.4) | 6.3(6.3) |
| (5) | | -24.5(-23.8) | -24.8(-24.1) | -23.0(-22.3) | -4.9(-4.9) | -5.3(-5.3) |
| (6) | | -7.6(-7.6) | -8.0(-8.0) | -5.9(-5.9) | -3.4(-3.5) | -3.7(-3.8) |

**Journal of Molecular Modeling**

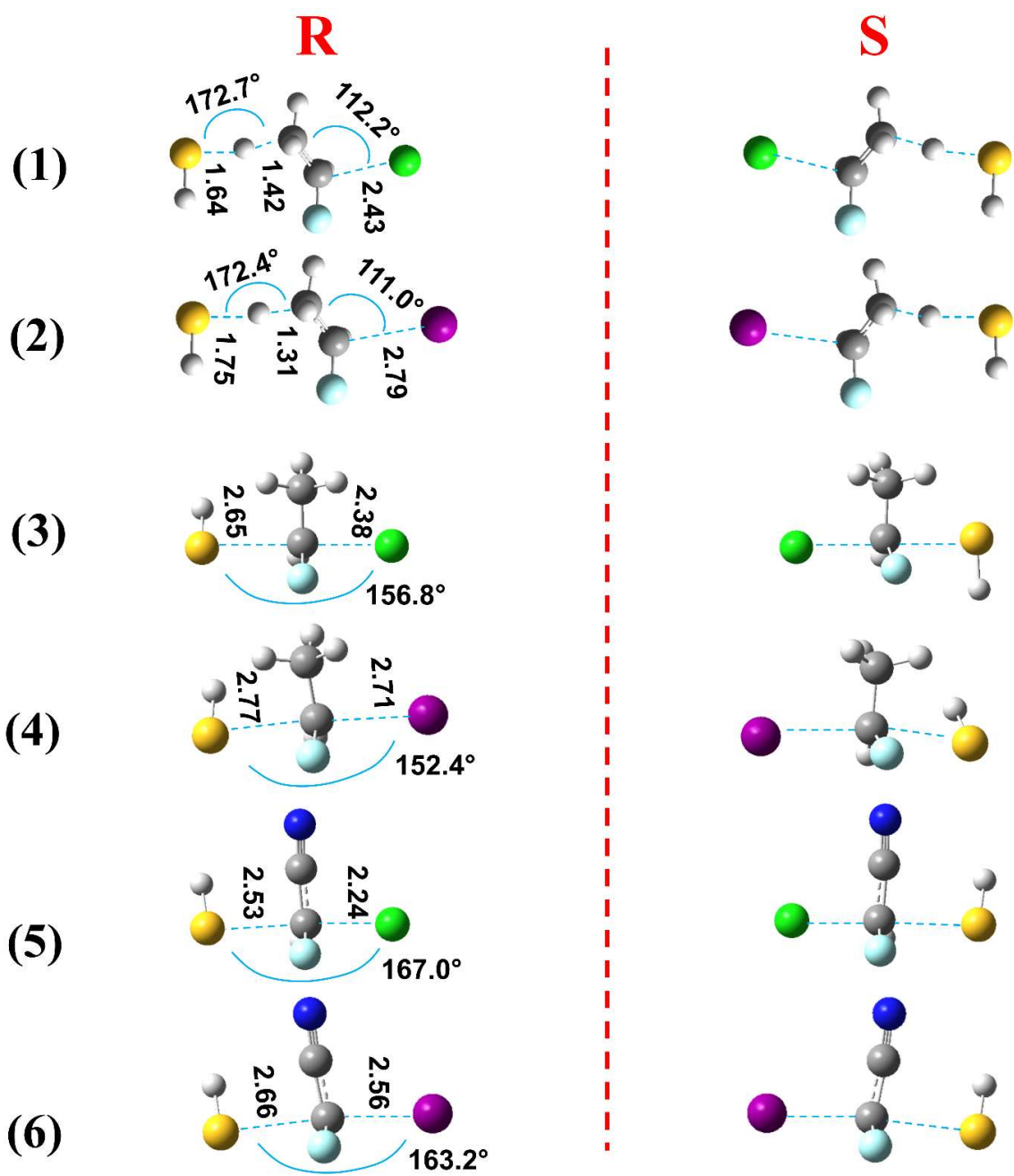

**Figure 1**

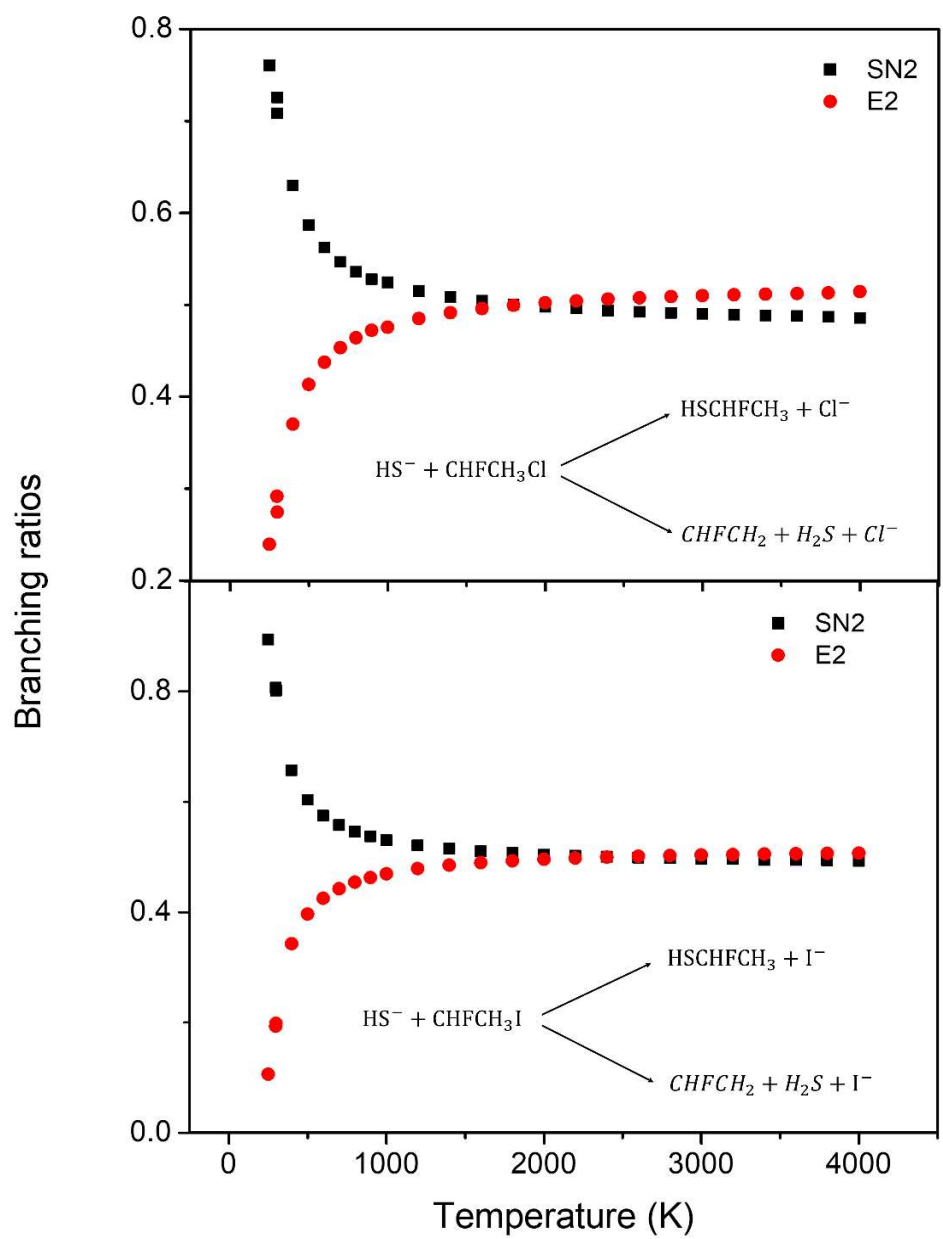

**Figure 2**

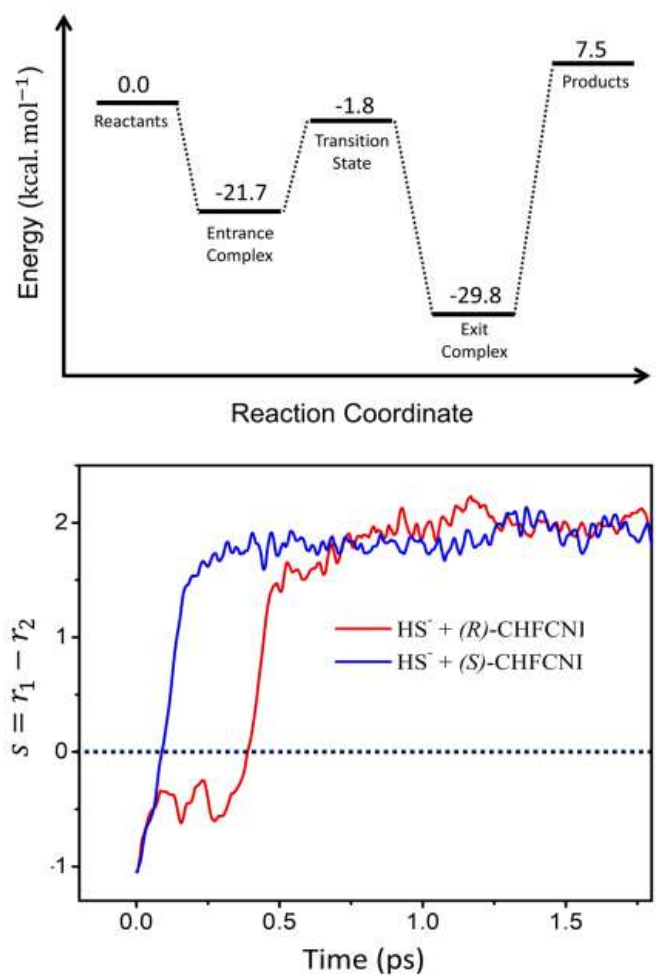

**Figure 3**